\documentclass[10pt,aps,prl,twocolumn,showpacs,superscriptaddress,nofootinbib,nobibnotes,longbibliography,floatfix]{revtex4-1}

\usepackage[utf8]{inputenc}
\usepackage[T1]{fontenc}
\usepackage{comment}
\usepackage{bm}
\usepackage{mathtools,amsmath,amssymb,amsfonts,mathrsfs,eucal,graphicx,tensor,csquotes,accents,commath,chngcntr,siunitx}
\usepackage[dvipsnames]{xcolor}
\usepackage[unicode]{hyperref}
\hypersetup{colorlinks=true, citecolor=MidnightBlue,
            linkcolor=MidnightBlue, urlcolor=MidnightBlue, linktocpage=true}
\usepackage[normalem]{ulem}

\DeclareMathAlphabet{\mathpzc}{OT1}{pzc}{m}{it}
\definecolor{darkgreen}{rgb}{0.0, 0.6, 0.0}

\begin{document}
\title{Exploring Ladder Symmetry and Love Numbers for Static and Rotating Black Holes}

\author{Chanchal Sharma}
\email{21510042@iitgn.ac.in }
\affiliation{Indian Institute of Technology, Gandhinagar, Gujarat 382055, India}

\author{Rajes~Ghosh}
\email{rajes.ghosh@iitgn.ac.in}
\affiliation{Indian Institute of Technology, Gandhinagar, Gujarat 382055, India}

\author{Sudipta~Sarkar}
\email{sudiptas@iitgn.ac.in}
\affiliation{Indian Institute of Technology, Gandhinagar, Gujarat 382055, India}

\begin{abstract}
Black hole solutions of general relativity exhibit a symmetry for the static perturbations around these spacetimes, known as ``ladder symmetry''. This symmetry proves useful in constructing a tower of solutions for perturbations and elucidating their general properties. Specifically, the presence of this symmetry leads to vanishing of the tidal love number associated with black holes. In this work, we find the most general spherical symmetric and static black hole spacetime that accommodates this ladder symmetry for scalar perturbation. Furthermore, we extend our calculations beyond spherical symmetry to find the class of stationary Konoplya-Rezzolla-Zhidenko black holes, which also possess a similar ladder structure.
\end{abstract}
\maketitle

\section{Introduction}
The concept of symmetry is arguably the most profound principle in physics. Symmetry principles often bear important theoretical consequences, which is greatly exemplified in the construction of both standard model of particle physics and theory of relativity. The notion of symmetry may also play further important role to find physics beyond these established theories.\\

In the present era of gravitational wave astronomy, when we are equipped with unprecedented technology to explore the features of extreme gravity, the implications of various symmetry principles might lead to far-reaching observational and theoretical consequences. A prime illustration of this is to understand the response of black holes (BHs) in an external tidal environment. The presence of a horizon imparts distinct characteristics to BHs in comparison to other astrophysical objects without horizons. Unlike such a horizonless compact  object, both Reissner-Nordstr\"{o}m and Kerr BH solutions in general relativity (GR) are known to have zero Love number \cite{Binnington:2009bb,Damour:2009vw,Kol:2011vg,Pani:2015hfa,Landry:2015zfa,LeTiec:2020spy,Chia:2020yla}, quantifying the vanishing tidal deformation under an external perturbation. This intriguing result can be interpreted as a manifestation of the celebrated no-hair theorems for BHs in GR~\cite{Bekenstein:1971hc, Bekenstein:1972ky, Gurlebeck:2015xpa}, establishing a natural connection between the presence of BH hairs and their tidal response.\\

In the conventional method for computing the tidal Love number (TLN), we study the linear perturbations around an asymptotically flat BH spacetime. The radial component of such a perturbation obeys a second-order differential equation, yielding two linearly independent solutions. At large distances away from the central BH, these solutions manifest as the tidal field growing as $r^\ell$, and the static response decaying as $r^{-\ell - 1}$. Here, the integer $\ell \geq 0$ represents the multipole order of the perturbing field. Then, motivated by an analogous Newtonian calculation~\cite{Binnington:2009bb, Damour:2009vw,Poisson:2014,Cardoso:2017cfl}, the TLN is defined as the ratio of the coefficient of the decaying tail to that of the tidal field. Utilizing this definition and considering the divergence of the static response at the horizon, it becomes evident that a Reissner-Nordstr\"{o}m/Kerr BH have zero Love number.\\ 

Apart from the aforesaid standard calculation, it has been recently demonstrated that the vanishing of TLN of BHs in GR can be attributed to a fundamental symmetry, known as the \textit{ladder symmetry} \cite{Hui:2021vcv,Berens:2022ebl,BenAchour:2022uqo,Katagiri:2022vyz,Hui:2022vbh}. As a consequence of this symmetry, the Hamiltonian corresponding to the scalar/vector/gravitational perturbations in both Reissner-Nordstr\"{o}m and Kerr background enjoys a decomposition in terms of the so-called raising and lowering operators analogous to that of a quantum harmonic oscillator. Then, vanishing of TLN follows directly by repeated application of the raising operator on the ``ground state'' solution, which has zero Love number.\\

Inspired by these interesting ideas, our aim is to conduct a comprehensive study of the ladder symmetries associated with general static and stationary BH spacetimes. In particular, we try to answer the following question: \textit{How the existence of such a ladder structure constraints the background BH spacetime?} For this purpose, we start with an arbitrary spherically symmetric static metric and find the form of the metric components from the imposition of generic ladder symmetry. To keep our analysis theory-agnostic, we focus solely on the scalar perturbations. This allows us to construct the most general static and spherically symmetric BH spacetime possessing a ladder structure for such scalar perturbations. It turns out such a metric must have a form given by,

\begin{equation}
    \mathrm{d}s^2  = -\frac{\Delta_b(r)}{h(r)}\, \mathrm{d}t^2 + \frac{h(r)}{\Delta_b(r)}\, \mathrm{d}r^2+ h(r)\, \mathrm{d}\Omega_{(2)}^2\, ,
\end{equation}
where $\Delta_b(r)= r^2 -c_2\, r + c_3$ with $(c_2,c_3)$ being some constants and $h(r)$ is an arbitrary radial function. The zeros of the function $\Delta_b (r)$ determine the location of the horizon. We also extend our analysis to find the most general metric within the so-called Konoplya-Rezzolla-Zhidenko class that has a similar ladder structure.\\

It is intriguing that the imposition of the ladder symmetry leads to such severe constraints on the form of the BH metric. The same symmetry can then be used to conclude the vanishing of the tidal Love number (TLN) associated with these black hole solutions. Though the existence of the ladder symmetry is not limited to black holes alone, it is important to note that the mere presence of ladder symmetry does not automatically ensure the vanishing of the TLN. In fact, to establish that TLN vanishes from ladder symmetry requires additionally the presence of a regular horizon. Specifically, this condition is crucial for demonstrating that the zeroth mode ($\ell=0$) of the perturbation have a zero TLN. Only then can one show that all higher modes also have zero TLN by a repeated application of the raising operator. Considering that astrophysical BHs are rarely isolated and are under constant external influence, our explorations may enhance the understanding of how BHs behave in the presence of perturbations---an aspect of central observational importance \cite{Fang:2005qq,DeLuca:2021ite, Creci:2021rkz,Cai:2019npx, Tan:2020hog, Charalambous:2023jgq, Katagiri:2023umb,DeLuca:2022xlz,Katagiri:2023yzm, Zi:2023pvl,Nair:2022xfm,Chakraborty:2023zed}.

\section{Review of Ladder Symmetries in Reissner-Nordstr\"{o}m and Kerr Case}
Before we move on to a more general calculation, it is useful to recall the computation of ladder symmetry in the Reissner-Nordstr\"{o}m and Kerr BH spacetimes. For our purpose, we shall only focus on the tidal response of these BHs in a scalar environment.\\

In the presence of a massless, static scalar field, the relevant perturbation equation takes the well-known Klein-Gordon form: $\Box\,\Phi(r, \theta, \phi) =0$. Here, the d'Alembertian operator is defined with respect to the background metric, which for a Reissner-Nordstr\"{o}m BH with mass $M$ and electric charge $Q$ ($M \geq |Q|$) is given by
\begin{equation}\label{RN metric}
   \mathrm{d} s^2 = - f_{RN}(r)\, \mathrm{d} t^2 + \frac{\mathrm{d} r^2}{f_{RN}(r)} + r^2\, \mathrm{d}\Omega_{(2)}^2\, ,
\end{equation}
where $f_{RN}(r)= 1-r_s/r+r_Q^2/r^2$ with $r_s = 2M$ and $r_Q = Q$. Hence, the horizons are located at $r_\pm = r_s/2 \pm [(r_s/2)^2-r_Q^2]^{1/2}$. Then, using a mode decomposition of the static scalar field in terms of spherical harmonics $\Phi_{\ell m} (r, \theta, \phi) = \phi_\ell(r) Y_{\ell m}(\theta,\phi)$, the Klein-Gordon equation can be reduced to a second-order radial differential equation: $\partial_r(\Delta\, \partial_r\phi_\ell)-\ell(\ell+1)\phi_\ell=0$,
where $\Delta(r) = r^2\, f_{RN}(r)$. This can be rewritten in a very suggestive form as $H_\ell\,\phi_\ell\,=0$, with the following definition of the Hamiltonian,
\begin{equation}\label{RN H}
    H_\ell= -\Delta(r) \left[\partial_r\big\{\Delta(r)\partial_r\big\}-\ell \big(\ell+1\big) \right]\, .
\end{equation}
In analogy to quantum harmonic oscillator, the above Hamiltonian then supports factorization in terms of two first-order operators~\cite{Hui:2021vcv,Berens:2022ebl},
\begin{equation}\label{RN D}
    D_\ell^+ = -\Delta(r)\,\partial_r - \frac{\ell+1}{2}\Delta'(r)\, ;\, 
    D_\ell^- = \Delta(r)\,\partial_r - \frac{\ell}{2}\Delta'(r)\, ,
\end{equation}
which are coined as the raising and lowering operators, respectively. As their names suggest, $D_\ell^\pm$ connects the radial solution $\phi_\ell$ to $\phi_{\ell \pm 1}$. In mathematical terms, this translates into two commutation relations with the Hamiltonian,
\begin{equation}\label{RN DD}
    H_{\ell+1}\, D_\ell^+ = D_\ell^+\,  H_\ell\, ;\,  H_{\ell-1}D_\ell^- = D_\ell^-H_\ell\, .
\end{equation}
Also, the Hamiltonian is related to the two ladder operators as
\begin{equation}
\begin{split}\label{RN HD}
    H_\ell &= D_{\ell-1}^+ D_\ell^- - \frac{\ell^2}{4}(r_s^2-4r_Q^2) \\ 
    & = D_{\ell+1}^- D_\ell^+ - \frac{(\ell+1)^2}{4} (r_s^2-4r_Q^2)\, .
\end{split}
\end{equation}
The above set of relations in Eq.~\eqref{RN DD} and Eq.~\eqref{RN HD} define a ladder structure, which plays a key role in showing the vanishing of Love numbers for Reissner-Nordstr\"{o}m BHs. For this purpose, let us first observe that for $\ell = 0$, $\phi_0 = \text{constant}$ represents an allowed solution to the radial equation $H_\ell \phi_\ell = 0$. Then, any other solution with higher $\ell > 0$ values can be constructed from $\phi_0$ by a repeated application of the raising operator as $\phi_\ell \propto D_{\ell-1}^+D_{\ell-2}^+...D_1^+D_0^+\phi_0$.
Note that such a solution yields a growing $r^\ell$ branch at infinity, which represents the tidal field in Newtonian terminology. Moreover, the ladder symmetry leads to a Noether current defined as $P_\ell (r) = \Delta\, \partial_r\left(D_1^- D_2^- \cdots D_\ell^- \phi_\ell \right)$, which is conserved $\partial_r P_\ell (r) = 0$ on-shell. The utility of this conserved quantity lies in understanding how the asymptotic solutions with particular behaviors get connected to the near horizon ones without explicitly solving the differential equation.\\

For example, it is easy to see that $\phi_\ell \propto D_{\ell-1}^+D_{\ell-2}^+...D_1^+D_0^+\phi_0$ for $\phi_0 = \text{constant}$ yields $P_\ell = 0$ and hence, the corresponding $\phi_\ell$ must be regular on the horizon. Apart from this regular solution, there is an independent decaying response proportional to $r^{-\ell - 1}$ at infinity, which leads to a non-zero $P_\ell$ at infinity. Then, the conservation of $P_\ell$ implies that this decaying solution at infinity must diverge as $\ln\left(r/r_s-1\right)$ near the horizon, which must be discarded. Hence, the Love number, specified by the ratio of the decaying tail to the growing one, should vanish identically. \\

A similar line of reasoning follows in the case of a rotating Kerr BH as well. For mass $M$ and spin angular momentum $a$ ($M \geq |a|$), the Kerr metric is given by
\begin{equation}
\begin{split}\label{Kerr metric}
    \mathrm{d} s^2=-\frac{\Delta}{\rho^2} & \left(\mathrm{~d} t-a\, \sin ^2 \theta\, \mathrm{d} \varphi\right)^2+\frac{\rho^2}{\Delta} \mathrm{d} r^2+\rho^2 \mathrm{~d} \theta^2 \\
    & +\frac{\sin ^2 \theta}{\rho^2}\left(a \mathrm{~d} t-\left(r^2+a^2\right) \mathrm{d} \varphi\right)^2
\end{split}
\end{equation}
where $\rho = r^2 + a^2 \cos^2\theta$ and $\Delta = r^2 -rr_s +a^2$. Thus, the inner/outer horizons are located at $r_\pm = r_s/2 \pm \sqrt{(r_s/2)^2-a^2}$. In such a spacetime, the scalar perturbation equation boils down to 
\begin{equation}\label{Kerr eom}
    \partial_r(\Delta\, \partial_r\phi_\ell)+\frac{a^2m^2}{\Delta}\phi_\ell-\ell(\ell+1)\phi_\ell=0.
\end{equation}
Then, following Ref.~\cite{Hui:2021vcv}, we can rewrite the above equation in a form analogous to the Reissner-Nordstr\"{o}m case:
\begin{equation}\label{Kerr H}
    H_\ell\,\phi_\ell\,=0, \,\, H_\ell= -\Delta\Big[\partial_r\left\{\Delta(r)\partial_r
    \right\}+\frac{a^2m^2}{\Delta}-\ell(\ell+1)\Big].
\end{equation}
A ladder structure resembling the one described for
the static spherically symmetric BH is present in
Kerr case too, which becomes apparent by defining the
ladder operators as
\begin{align}\label{Kerr D}
    & D_\ell^+ = -\Delta\,\partial_r + \frac{\ell+1}{2}(r_s-2r)\, ,\\ \nonumber
    & D_\ell^- = \Delta\,\partial_r + \frac{\ell}{2}(r_s-2r).
\end{align}
These operators follow relations similar to Eq.~\eqref{RN DD}. The behavior of Eq.~\eqref{Kerr H} at the two asymptotes can be examined here as well. At large $r$, the two independent solutions for $\phi_\ell$ are $r^\ell$ and $r^{-\ell - 1}$. Whereas in the near-horizon limit $z \rightarrow z_k$, $\phi_\ell$ goes as either $\text{constant}$ or as $e^{-2 i q \ln(z/z_k-1)}$, where $q = a\, m/ z_k$ with $z=r-r_-$ and $z_k = r_+-r_-$. Among them, the former is regular at the horizon and can be raised to the solution with multipole $\ell$ using $\phi_\ell = D_{\ell-1}^+D_{\ell-2}^+...D_1^+D_0^+\phi_0$. This implies, $\phi_\ell \sim 1 + z+...+z^\ell$ manifests itself as a polynomial with no decaying behavior. Moreover, following Ref.~\cite{Hui:2021vcv}, one may construct an analogous Noether current $P_\ell(r)$ in the Kerr case also, which implies the other decaying solution at infinity must diverge at $r_+$. Combining these two facts, it is evident that scalar TLN vanishes for Kerr BHs as well.\\

In Refs.~\cite{Charalambous:2021mea,Hui:2021vcv,Hui:2020xxx,Katagiri:2022vyz}, the authors have further shown the presence of a ladder symmetry among different spin-perturbations, namely the scalar ($s=0$), vector ($s=1$), and gravitational ($s=2$) ones. We encourage our readers to follow this nice construction, which demonstrates why the Reissner-Nordstr\"{o}m/Kerr BHs have vanishing Love numbers even for static higher-spin perturbations. However, for the purpose of this paper, we shall skip those computations.\\

It is important to note that the structure of the ladder operators $D_\ell^{\pm}$ and the ladder symmetry are closely tied to the particular form of the Hamiltonian operator $H_\ell$, which in turn depends on the background metric. Thus, such ladder structure is not generally expected to hold for an arbitrarily spacetime. As an example, suppose we consider a theory of gravity with higher curvature terms, then its solution may deviate from Reissner-Nordstr\"{o}m and Kerr metrics in such a way that does not support ladder structure for scalar perturbations. This motivates us to find the most general class of static and stationary spacetimes which admit such ladder symmetry.

\section{Generalization for Static Spherically Symmetric BH}
We shall now shift our attention to study the tidal response of a general static and spherically symmetric metric under the influence of a static and massless scalar field. Our goal is to derive the constraints on the form of such a metric by demanding the existence of a ladder structure. A generic static, spherically symmetric metric is given by
\begin{equation}\label{SSS gen metric}
    \mathrm{d}s^2 = -f(r) \mathrm{d}t^2 + \frac{\mathrm{d} r^2}{g(r)} + h(r) \mathrm{d}\Omega_{(2)}^2\, .
\end{equation}
In such a spacetime, the massless Klein-Gordon equation can always be reduced to the form: $\Delta_b(r)\, \phi_\ell''(r)+\Delta_c(r)\, \phi_\ell'(r)-\ell(\ell+1)\phi_\ell(r) = 0$. Here, the explicit forms of $\{\Delta_b,\, \Delta_c\}$ depend on the components of the background metric. Motivated by the Reissner-Nordstr\"{o}m case presented earlier, it is suggestive to multiply the above equation by $\Delta_b$ and define the general Hamiltonian as 
\begin{equation}\label{SSS H}
    H_\ell=-\Delta_b(r)\;\Big[\Delta_b(r)\, \partial_r^2+\Delta_c(r)\, \partial_r-\ell(\ell+1)\Big]\, .
\end{equation}
As before, this multiplicative factor makes the Hamiltonian nicely factorizable, and all the subsequent expressions look cleaner.\\

We want to derive conditions on $\{\Delta_b,\, \Delta_c\}$ so that the quadratic Hamiltonian given by Eq.~\eqref{SSS H} supports a ladder structure similar to Eq.~\eqref{RN DD} and Eq.~\eqref{RN HD}. For this purpose, our first step is to decompose this $H_\ell$ into two first-order raising and lowering operators. Taking inspiration from the structure of ladder operators for the Reissner-Nordstr\"{o}m case, we define them as
\begin{equation}
\begin{split}\label{SSS RL}
    &D_\ell^+ = -\Delta_1(r) \partial_r+\frac{\ell+1}{2}\Delta_2(r)\, ; \\
    &D_\ell^- = \Delta_3(r) \partial_r+\frac{\ell}{2}\Delta_4(r)\, .
    \end{split}
\end{equation}
So far these $\Delta$'s are some unknown functions of $r$ only, independent of $\ell$. To determine their functional forms, we now employ the fundamental commutation relations given by Eq.~\eqref{RN DD}, which they must satisfy with the Hamiltonian for an arbitrary choice of $\phi_\ell$. Therefore, we get the following conditions,
\begin{equation}
\begin{split}\label{SSS D12}
    &\Delta_1(r)=\Delta_b(r)\;,\;\Delta_2'(r)= -2-\frac{s'(r)}{\ell+1}\, ; \\ 
    & \Delta_3(r)=\Delta_b(r) \;,\;\Delta_4'(r)=-2+\frac{s'(r)}{\ell}\, ;\\
    & \Delta_{b}'(r)= - \Delta_{2}(r)-\frac{s(r)}{\ell+1}=- \Delta_{4}(r)+\frac{s(r)}{\ell}\, ,
\end{split}
\end{equation}
where $s(r) = \Delta_c(r)-\Delta_b'(r)$. Using the fact that $\Delta_2$ and $\Delta_4$ do not depend on $\ell$, we must set $s(r) = c_1$ with $c_1$ being an $\ell$-independent constant. Then, using the last relation along with the fact that $\Delta_b$ is $\ell$-independent, we get $c_1 = 0$. As a result, $\Delta_2(r) = \Delta_4(r) = -2\, r + c_2$, and 
\begin{equation}\label{SSS Del}
    \Delta_b(r)= r^2 -c_2\, r + c_3\, ,
\end{equation}
where $(c_2, c_3)$ are again some $\ell$-independent constants. It is remarkable that the imposition of the ladder symmetries led to such a simple forms of various functions like $\Delta_{b}(r)$. Moreover, the Hamiltonian takes the form similar to Eq.~\eqref{RN H},
\begin{equation}\label{SSS Hn}
    H_\ell= -\Delta_b(r) \left[\partial_r\big\{\Delta_b(r)\partial_r\big\}-\ell \big(\ell+1\big) \right]\, .
\end{equation}
Let us now summarize the ladder structure we have obtained so far, 
\begin{equation}
    \begin{split}\label{SSS HRL}
    &D_\ell^+ = -\Delta_b(r)\,\partial_r - \frac{\ell+1}{2}\, \Delta'_b(r)\, ,\\ 
    &D_\ell^- = \Delta_b(r)\,\partial_r - \frac{\ell}{2}\, \Delta'_b(r)\, ,\\
    &H_\ell = D_{\ell+1}^- D_\ell^+ - \frac{(\ell+1)^2}{4}\left(c_2^2-4c_3\right) \\
    &\, \, \, \, \, \, \, \, = D_{\ell-1}^+ D_\ell^- - \frac{\ell^2}{4}\left(c_2^2-4c_3\right)\, .
\end{split}
\end{equation}
Additionally, the form of the Hamiltonian given by Eq.~\eqref{SSS Hn} along with Eq.~\eqref{SSS Del} also requires that the metric components in Eq.~\eqref{SSS gen metric} must satisfy the relation
\begin{equation}\label{f(r)}
    f(r)= g(r) = \frac{\Delta_b(r)}{h(r)}\, .
\end{equation}
We obtain this relation by comparing the Klein-Gordon Hamiltonian obtained from Eq.~\eqref{SSS gen metric} with that in Eq.~\eqref{SSS Hn}. \\

\textit{Therefore, the most general static and spherically symmetric metric that supports the ladder symmetry can be written as}
\begin{equation}\label{SSS metric}
    \mathrm{d}s^2  = -\frac{\Delta_b(r)}{h(r)}\, \mathrm{d}t^2 + \frac{h(r)}{\Delta_b(r)}\, \mathrm{d}r^2+ h(r)\, \mathrm{d}\Omega_{(2)}^2\, ,
\end{equation}
where $\Delta_b(r)$ is given by Eq.~\eqref{SSS Del}. The above metric is spatially conformal to Reissner-Nordstr\"{o}m metric with a conformal factor $h(r)/r^2$, provided we identify $c_2 \to r_s$ and $c_3 \to r_Q^2$. Some comments on the above construction are in order:\\
\\
(i) Note that the metric in Eq.~\eqref{SSS gen metric} did not have the property $g_{tt}\, g_{rr} = -1$ to begin with. However, the imposition of the ladder symmetry has forced this structure in Eq.~\eqref{f(r)}. Hence, if we insist that the metric in Eq.~\eqref{SSS metric} is a solution of GR with some matter $T_{\mu \nu}$, then it must have vanishing radial null-null component \cite{Jacobson}, i.e., $T_{\mu \nu}\, k^\mu\, k^\nu = 0$ with $k^\mu$ being the radial null vector. For example, it is easy to check that this condition is satisfied both in vacuum and electro-vacuum.\\
 \\
(ii) Since the function $h(r)$ remains unconstrained, we can maintain the ladder structure by choosing it at our will (as long as it does not produce any singularity in the domain of outer communication). For example, the choice of $h(r)=r^2$ leads to a Reissner-Nordstr\"{o}m-type metric. Had we fixed $f(r) = r^2$ from the beginning, we would have missed this additional freedom.\\ 
\\
(iii) However, for generic choices of $h(r)$, the metric will not (in general) be diffeomorphic to Reissner-Nordstr\"{o}m, which can be readily checked by calculating various curvature scalars. For the purpose of illustration, let us consider the Ricci scalar,
\[
    R=\left[ h'(r)^2-2\, h(r)\, h''(r) \right]\, \frac{\Delta_b(r)}{2\, h^3(r)}\, .
\]
In contrast to the $4$-dimensional Reissner-Nordstr\"{o}m metric, $R$ does not vanish unless $h(r) = (a\, r + b)^2$ with $(a, b)$ being some constants. \\
\\
(iv) The ladder structure does not determine the sign of the constants $(c_2,c_3)$ appearing in the metric in Eq.~\eqref{SSS metric}. Though if we further require that the associated spacetime is that of a BH, then we must impose $c_2^2 \geq 4\, c_3$. This would ensure the existence of a positive real root of $\Delta_b(r) = 0$ at $r = c_2/2 + \sqrt{c_2^2/4-c_3}$\, .\\
\\
(v) Moreover, since the metric in Eq.~\eqref{SSS metric} gives rise to a scalar Hamiltonian similar to the Reissner-Nordtr\"{o}m BH (assuming $c_2^2 \geq 4\, c_3$), one can easily follow the Love number calculation presented earlier to show that these BHs also have zero TLN.

\subsection{Generalization for Rotating BH: Konoplya-Rezzolla-Zhidenko Class}
In this section, we aim to extend our previous result beyond spherical symmetry. Ideally, one would like to find the most general stationary BH solution which admits the ladder symmetry for static massless scalar perturbations. However, in such a background, the Klein-Gordon equation will not in general be separable in a radial and angular parts. To avoid this difficulty, we instead start with a well-motivated generalization of the Kerr-like spacetimes, namely the Konoplya-Rezzolla-Zhidenko (KRZ) class\footnote{It extends the previously proposed framework of Ref.~\cite{Rezzolla:2014mua} for static and spherically symmetric metric in the presence of rotation.} of metrics~\cite{Konoplya:2016jvv, Konoplya:2018arm}. These BH spacetimes represents the most general stationary, axisymmetric and asymptotically flat Kerr-like spacetimes which admits the separation of the scalar wave equation into radial and angular parts in Boyer-Lindquist type coordinates. Such a metric can be written as~\cite{Konoplya:2018arm},
\begin{equation}
    \begin{split}\label{RZ metric}
    \mathrm{d}s^2 = &- \left(\frac{N^2}{K^2}-\frac{a^2R_M^2(1-y^2)}{r^4\Sigma^2K^2}\right) \mathrm{d}t^2+ K^2\, r^2 (1-y^2) \mathrm{d}\varphi^2\\
    &-\frac{2aR_M}{r\Sigma}\, (1-y^2)\, \mathrm{d}t \mathrm{d}\varphi
    +\Sigma\left(\frac{R_B^2}{N^2} \mathrm{d}r^2 +\frac{r^2\, \mathrm{d} y^2}{1-y^2}\right),
\end{split}
\end{equation}
with $y = \cos \theta$ as one of the coordinates, and the separability requires that the functions has following forms (where $a$ is the rotation parameter)
\begin{equation}
    \begin{split}\label{RZ metric P}
    &\Sigma(r,y) = R_\Sigma+\frac{a^2y^2}{r^2},N^2(r)=R_\Sigma-\frac{R_M}{r}+\frac{a^2}{r^2}, \\ 
    &K^2(r,y)=\frac{1}{\Sigma}\left[R_\Sigma^2+R_\Sigma\frac{a^2}{r^2}
    +\frac{a^2R_M}{r^3}\right] +\frac{a^2y^2N^2}{r^2\Sigma}.
\end{split}
\end{equation}
The location of the event horizon (which is also the Killing horizon) is given by $N(r) = 0$. Thus, the metric solely depends on the three functions of radial coordinates $R_\Sigma(r),\, R_M(r)$, and $R_B(r)$, out of which one can be fixed as per the gauge freedom. We choose $R_B(r) =1$ and the other two are independent functions of $r$ \cite{Konoplya:2018arm}. Moreover, the asymptotic flatness is assured, if $R_\Sigma(r)$ approaches unity and $\frac{R_M(r)}{r^2}$ vanishes in $r\to\infty$ limit.\\

Then, in the background of such a metric, the radial part of the Klein-Gordon equation for a static scalar field $\Phi_{\ell m} = \phi_\ell(r) Y_{\ell m}(y, \varphi)$ simplifies to
\begin{equation}\label{RZ H}
    H_\ell\,\phi_\ell\,=-\Delta\left[\partial_r\left\{\Delta(r)\partial_r\right\}+\delta(r)-\ell(\ell+1)\right]\phi_\ell\, = 0\, ,
\end{equation}
where $H_\ell$ is the Hamiltonian, $\Delta$ and $\delta$ are given by
\begin{equation}\label{RZ Del}
\Delta(r) = r^2 N^2(r) = r^2 R_\Sigma(r) - r R_M(r) +a^2\:,\:\delta(r)=\frac{a^2m^2}{\Delta(r)}\, ,
\end{equation}
with $m$ being the azimuthal number. We aim to study this scalar field equation in the KRZ class of BH spacetimes and investigate the existence of the ladder structure. We shall show that the requirement of the ladder symmetry would fix the functional form of $\Delta$. For this purpose, we define the raising and lowering operators as
\begin{equation}\label{RZ D}
    D_\ell^+ = -\Delta_1\,\partial_r + \frac{\ell+1}{2}\,\Delta_2\,;\,D_\ell^- = \Delta_3\,\partial_r + \frac{\ell}{2}\,\Delta_4.
\end{equation}
where $\Delta$'s could in principle depend on not only $r$ but $\ell$ due to the absence of spherical symmetry. Substituting $D_\ell^+$ in the fundamental commutation relation given by Eq.~\eqref{RN DD}, we obtain
\begin{equation}\label{RZ 12}
    \begin{split}
        &\Delta_1(r,\ell)= \Delta(r) \;,\; \Delta_2(r,\ell)= -2 r + e_2, \\
        &\delta(r)= \frac{e_4}{\Delta(r)}-\frac{(\ell+1)^2\, (r^2-\Delta-e_2\, r+e_3)}{\Delta(r)}\, .
    \end{split}
\end{equation}
Comparing with Eq.~\eqref{RZ 12}, one further gets $e_4 = a^2\, m^2$, and 
\begin{equation}\label{RZ Delta}
    \Delta(r) = r^2-e_2\, r+e_3\, .
\end{equation}
Hence, the constants $(e_2, e_3, e_4)$ are also independent of $\ell$. Similarly, for the lowering operator $D_\ell^-$, the fundamental commutation relation in Eq.~\eqref{RN DD} gives
\begin{equation}\label{RZ 34}
    \begin{split}
        &\Delta_3(r,\ell)=\Delta(r)\;,\; \Delta_4(r,\ell)=-2r+e_5, \\
        &\Delta(r)= e_6+\frac{(2\ell+1)r^2-e_2(\ell+1)^2\, r+e_5\ell^2\, r}{2\ell+1}\, .
    \end{split}
\end{equation}
Comparing this form of $\Delta$ with that given in Eq.~\eqref{RZ Delta}, we obtain $e_5 = e_2$ and $e_6 = e_3$. Then, it is easy to check that $H_\ell$ can be factorized as
\begin{equation}\label{RZ HRL}
    \begin{split}
        &H_\ell=\;D_{\ell+1}^- D_\ell^+-\frac{(\ell+1)^2}{4}\left(e_2^2-4e_3\right)-a^2\, m^2 \\
        &\, \, \, \, \, \, \, \,=D_{\ell-1}^+ D_\ell^- - \frac{\ell^2}{4}\left(e_2^2-4e_3\right)-a^2\, m^2\, .
    \end{split}
\end{equation}
Lastly, it remains to find the form of $R_\Sigma(r)$ and $R_M(r)$ appearing in the metric given by Eq.~\eqref{RZ metric}. This can be achieved by comparing the functional form of $\Delta(r)$ in Eq.~\eqref{RZ Del} and Eq.~\eqref{RZ Delta}. Specifically, the coefficients of $r^2$, $r$, and the constant term in Eq.~\eqref{RZ Del} should be $1$, $-e_2$, and $e_3$, respectively. \\

A simpler illustration could be to choose one of the radial functions as a constant, and find the corresponding class of metrics. If we set $R_{M}(r) = e_0$ (constant), we would get the second function as,

\begin{equation}\label{RZ RM}
    R_\Sigma(r) = 1+\frac{b}{r}+\frac{d}{r^2}\, ,
\end{equation}
with $b= e_0 -  e_2 \, ,\; \textrm{and}\; d = e_3 - a^2 \, .$ Note that both Kerr and Kerr-Sen spacetimes are members of this class \cite{Konoplya:2018arm}. Also, it is easy to check that in the limit of $a\rightarrow0$, the metric in Eq.~\eqref{RZ metric} (along with Eq.~\eqref{RZ Delta}) reduces to the spherically symmetric spacetime given by Eq.~\eqref{SSS metric}, provided we make the identification $\Delta_{b}(r) = r^2\, N^2 = \Delta(r)$ and $h(r) = r^2\, \Sigma (r)$.\\

Moreover, since the metric in Eq.~\eqref{RZ metric} gives rise to a scalar Hamiltonian similar to the Kerr BH (assuming $e_2^2 \geq 4\, e_3$), one can easily follow the Love number calculation presented earlier to show that these BHs also have zero TLN.

\section{Conclusion and Discussions}
Unlike horizonless compact objects, the Reissner-Nordstr\"{o}m and Kerr BHs exhibit zero TLN, quantifying the vanishing tidal deformation under an external perturbation. Consequently, any nonzero values of TLN would indicate deviation from such spacetime geometries~\cite{Cai:2019npx, Tan:2020hog, Charalambous:2023jgq, Katagiri:2023umb} and/or departure from the classical BH paradigm~\cite{Nair:2022xfm, Cardoso:2017cfl, Chakraborty:2023zed}. Both of these possibilities are well-studied in literature as they provide us with a powerful observational tool to probe such possibilities~\cite{DeLuca:2022xlz, Katagiri:2023yzm, Zi:2023pvl}.\\

However, one faces two major difficulties in the traditional way of calculating TLN. Firstly, apart from GR, the Teukolsky-like equation for gravitational perturbations in most of the modified theories is not known. Secondly, even for scalar perturbations that does not require any field equations, a case-by-case study of TLN for all possible metrics is highly tedious and inefficient. In this context, other tools such as the notion of ladder symmetry provide us with a unified and efficient way to infer the Love number. Interestingly, both the Reissner-Nordstr\"{o}m and Kerr BHs support such ladder structure for static scalar (also vector and gravitational) perturbations, indicating vanishing of TLN. \\

Motivated by this important result, we have presented the most general static and stationary (in KRZ class) BH metrics having ladder symmetry for static (with frequency $\omega = 0$) scalar perturbations. This in turn implies that the corresponding BHs have zero TLN for $\omega = 0$, i.e., $\Lambda = \mathcal{O}(M\, \omega)$. Actually, for the case of static BHs, our result is even stronger. In particular, since $H_\ell\,\phi_\ell \propto \omega^2$ for non-static perturbations, the corresponding TLN must be $\mathcal{O}(M^2\omega^2)$. However, a similar assertion does not hold for the rotating case, because $H_\ell\,\phi_\ell \propto \omega$ for non-static perturbations. These conclusions match with the recent claim reported in Ref.~\cite{Perry:2023wmm}. We have also found that for these background spacetimes, the ladder symmetry is also a symmetry at the level of the action of these perturbing fields.\\

Given the immense theoretical and observational significance, it will be interesting to extend our work for gravitational perturbations as well as for non-liner perturbations \cite{DeLuca:2023mio}. Another important prospect would be studying various properties like geodesic structure, shadow, and stability of the general class of BH spacetimes given by Eq.~\eqref{SSS metric} and Eq.~\eqref{RZ metric}. We leave these analyses for a future attempt. 

\section{Acknowledgement}
We thank the anonymous referee for useful comments on our work. CS acknowledges the support from the Sabarmati Bridge Fellowship (Project ID: MIS/IITGN-SBF/PHY/SS/2023-24/025) from IIT Gandhinagar. The research of R.G. is supported by the Prime Minister Research Fellowship (PMRF ID: 1700531), Government of India. S.~S. acknowledges support from the Department of Science and Technology, Government of India under the SERB CRG Grant (CRG/2020/004562).

\end{document}